# Designing multi-model conversational AI financial systems: understanding sensitive values of women entrepreneurs in Brazil

HELOISA CANDELLO, IBM Research
GABRIEL MENEGUELLI SOELLA, IBM Research, University of São Paulo
LEANDRO DE CARVALHO NASCIMENTO, Assis Chateaubriand Foundation

Small business owners (SBOs), specially women, face several challenges in everyday life, especially when asking for microcredit loans from financial institutions. Usual difficulties include low credit scores, unbaked situations, outstanding debts, informal employment situations, inability to showcase their payable capacity, and lack of financial guarantor. Moreover, SBOs often need help applying for microcredit loans due to the lack of information on how to proceed. The task of asking for a loan is a complex practice, and asymmetric power relationships might emerge, but that benefits micro-entrepreneurs only sometimes. In this paper, we interviewed 20 women entrepreneurs living in a low-income community in Brazil. We wanted to unveil value tensions derived from this practice that might influence the design of AI technologies for the public. In doing so, we used a conversational system as a probe to understand the opportunities for empowering their practices with the support of AI multimedia conversational systems. We derived seven recommendations for designing AI systems for evaluating micro-business health in low-income communities.

**ACM Reference Format:**
Heloisa Candello, Gabriel Meneguelli Soella, and Leandro de Carvalho Nascimento. 2024. Designing multi-model conversational AI financial systems: understanding sensitive values of women entrepreneurs in Brazil. In *Proceedings of the 2024 ACM International Conference on Interactive Media Experiences Workshops (IMXw '24), June 12, 2024, Stockholm, Sweden.* ACM, New York, NY, USA, 11 pages. https://doi.org/10.1145/3672406.3672409

## 1 INTRODUCTION

In this paper we aim to explore the nuances and values of designing multimedia conversational user interfaces for women entrepreneurs in vulnerable situations. Particularly, key research questions guided our research - **[1] What are the perceived inherent values, and value tensions built in the everyday practices of credit access in low-income communities? [2] What values should be considered when designing conversational user interfaces (CUI) to evaluate micro-business health and promote credit access in low-income communities?** Designing technology for vulnerable communities should consider the main values built into their everyday practices. Women face several social and stigma challenges when applying for loans due to informal settings of work, lack of credit score and proof of their revenues. Additionally, the lack of financial education and the emergency of their everyday situations might affect their financial planning, and future outcomes. This reality makes difficult to access the health and status of their businesses using formal financial institutions criteria. In this research we interviewed 20 women entrepreneurs, living in low-income communities applying the Value-sensitive design approach [23] to unveil the main sensitive-values that should be considered when designing multi-model conversational AI financial systems for this audience. The contributions of this research to IMX community relies on: *(1) A set of values, and value tensions to consider when using AI models to design multimedia conversational systems for vulnerable communities. (2) Design recommendations on how multi-model conversational systems based on artificial intelligence can be employed to mitigate, and sometimes amplify those relationships.*

## 2  MICROCREDIT PRACTICES

In this paper, we focus on the social and economic context of microcredit, which can be defined as a term that *"describes lending to poorer sections of the population and micro-enterprises, usually without material security for the loan."* [38].

Previous work exposed that digital financialization, provided by fintechs [5, 44] or mobile money apps [5, 34, 44, 48] are prominent in the Global South [26]. This process occurs among people with traditional bank accounts, but with the aim to promote financial inclusion of unbanked 2.2 billion people worldwide [48]. Although, some of the current digital financial services were identified as having unequal access [5] to promote financial inclusion in the Global South [5, 26, 38, 44]. In this context, self-entrepreneur women are the ones most impacted by inequality in financial services access. [25, 36, 38]. Additionally, the use of financial apps to access financial services is in expansion. [44]. The AI has been applied by fintechs to promote digital financial inclusion include "audio processing, knowledge representation, speech to text, deep learning, expert systems, natural language processing, machine learning (ML), robotics, and symbolic logic". And it is also evident that traditional banks adopted AI conversational systems to interact with customers. Those initiatives brought discussions of the benefits and harms, bias, and others vulnerabilities of the digital uses [10, 14, 27, 32, 33, 39, 40]. Moreover, this discussion is an emergent topic by researchers of Human-Computer Interaction, particularly using voice and text based conversational user interfaces (CUIs) as a platform for financial services. [30, 46, 47]. Financial institutions have traditional metrics to define the Creditworthiness of small business owners, for instance credit score metrics. In some cases, the values biasing credit scoring metrics might be understood as oppressive and stereotype indebted people named "delinquent" [9], highlighting the lack of understanding of social-contextual conditions. In financial initiatives using AI, it is essential to consider trustworthy factors, bias and implicit values to promote financial inclusion. Alternative criteria and values based on the social reality of people are essential to make possible sustainable and inclusive microcredit offers. Many approaches are adopted by researchers to understand social values in context, such as the studies applying Value Sensitive Design [4, 6, 20, 30, 41, 47]. In order to unveil values in real-world situations, for instance: fairness [2, 18, 21, 30, 37], explainability [18, 19, 41], accountability [1, 16, 19, 37], trust [2, 11, 16, 24, 45, 50], and privacy [29, 42, 43, 49].

## 3  METHODOLOGY

In this paper, we employed the theoretically grounded approach Value-sensitive design (VSD) [20, 22, 41] to support the design of a conversational system for micro-business owners in low-income communities considering the qualitative informed ethical views of small-business owners. In the process, we applied empirical-inductive methods [17] to unveil the main values and values tensions in the microcredit access activity. A reflexive thematic analysis was conducted for analysing the data [7, 8, 15]. Our Conceptual, empirical and technical investigations consist of understanding the main practices and values of small business owners, and potential microcredit customers. We conducted semi-structured interviews to understand the aligned values and consequent relationship tensions occurred in the credit access process. Value is defined here as *"what a person or group of people consider important in life"* [23].

We explored the views of microcredit customers in relation to their business practices and Artificial Intelligence perceptions. Participants were women enrolled in a financial education course provided by a NGO in low-income community in Brazil, mobile phones and services on WhatsApp were the main main tech used by them, twenty accepted to be part of the study. Most of them (12) worked with more than one product/service, whether manufactured by them or purchased for resale. Their activities were in beauty services, cake baking, handmade accessories (e.g. handbags, wallets, scarfs), perfumery/cosmetics and personalized gifts (table 1). The interviews were completed in 3 cycles. First, interviews were conducted virtually directly with researcher. The second cycle was in-person and had the researcher and a NGO representative conducting the interviews. The third cycle was in-person mediated by the same NGO representative.

The three cycles followed the same protocol. Interviews lasted 60 min. Participants reviewed a digital consent form explaining the study, and agreed to be recorded and photographed. We prepared a deck with photos to set up the pace of each stage of the interview and cover specific topics. One topic for each picture, and we showed the participants during the interviews. This approach was built based on [28] and [13]. The topics included descriptions of their every day practices, financial status, business management, financial tools, financial planning, experience with micro-credit access, and evaluating a concept of a multimedia chatbot.

In the first cycle participants watched a video-prototype picturing the design concept. In the second and third cycle, participants interacted with a multimedia conversational system pilot, a chatbot, similar to the one in the video prototype on WhatsApp. The chatbot displayed a health business index (HBI) based on the questions participants answered. Participants in cycle 1 saw the HBI in the video prototype and participants in cycle 2 and 3 experienced it in the interactive chatbot experience. (Fig 1). All them were females. Twelve participants informed that had previous interaction with chatbots mainly mediated by customer service from Banks, cell phones companies, and internet providers. Table 2 shows their main business activities.

## 4 DATA ANALYSIS

The data collected was transcribed and analysed by two of the authors of this paper. We followed a Thematic Analysis [8] methodology, identifying codes and themes constructing meaning through the coding process using an interpretative approach. We started coding process doing discussion sessions among the authors of this paper to unveil possible trends in the data. To facilitate the discussion, we used as seed codes the list proposed by [22] that were often implied in system design. Researchers conducted an open-coding analysis using the result of the team discussion as initial point to explore the data. The coders also highlighted the transcription excerpts describing the process of acquiring a loan, the role of each stakeholder mentioned in the process, challenges micro businesses owners face, technology adoption and Artificial Intelligence perceptions reported by participants. Both coders analysed the transcriptions alone, and had three follow-up meetings to engage in discussions regarding coding divergences to increase consistency and avoid individual biases of the researchers (R1 and R2), allowing them to arrive at consensus about the codes and themes in the data to proceed the analysis.

We followed the approach of [8] to establish inter-coder reliability while constructing meaning throughout the coding process. We adopted an inductive-iterative strategy to discuss the codes and themes and grasp the analytical views of the three researchers. We performed a "consensus coding" approach [31], in this approach coders engage in ongoing discussions regarding coding discrepancies to increase consistency and clarify individual biases of the three researchers, allowing them to reach consensus of the codes and themes in the data.

Both coders validated and agreed with 20 basic themes that were representative of the sensitive values in the process of assessing Creditworthiness. Both coders used this set of 20 codes to analyse the transcriptions that were missing.

The 20 codes were: Accountability, Autonomy, Coercion, Collaboration, Creditworthiness, Explainability, Fairness and Justice, Frustration, Human welfare, Identity, Legitimization, Mistrust, Ownership and property, Power dynamics, Prejudice, Privacy, Transparency, Trust, Universal Usability, Vulnerability. During the coding process, authors grouped those basic themes into seven global themes, based on their strong relation with other basic themes, and on their interpretative discussions using affinity diagrams [35]. The seven global themes were named as value dimensions: Creditworthiness, Accountability, Vulnerability, Fairness and Justice, Universal Usability, Ownership and Property, and Identity. Follow a brief description and the value dimensions in Table 1 (Appendix).

## 5 RESULTS

In this session, we discuss the seven value dimensions that emerged from the analysis of the real-world situations reported by the interviewed small business women. Those value-based design recommendations for conversational systems are displayed in table 2.

*5.0.1 Creditworthiness.* We define *Creditworthiness* as SBOs' capacity to pay bills on time, which is more than just a financial situation; this value affects participants' honor and reputation in their community. A vulnerable context of indebtedness is identified in the words of P20 *"Very important is to have name clean too. My [name] is not."* demonstrating *Accountability, Identity* and *Transparency* to explain her financial context. A similar debt scenario was perceived in P13 and P17. Additionally, self-organization and lack of financial literacy in financial planning (P13, P15) were essential for keeping the business health. For instance, financial planning learned in the financial education course can be a way to manage payment debts (P17) or a way to improve business management, as exposed by P15 - *"She [a financial educator] gave a suggestion [for me] to do a little spreadsheet for my business finance, I did, and I inserted my data on a spreadsheet, and for now I could see that from first month, I received 900 BRL. Well, not as profit, but all that money, 900 BRL. And I did not know I could receive this before adding everything to the spreadsheet.".* In low-income communities access to credit was observed as an informal practice, called the *"fiado"*, e. g., known as store credit, "put on the tab" situations [12]. It is based on collaboration, social capital, and community trust. SBOs also mentioned that *"fiado"* practice promotes a lack of cash when the debt is not paid (P6), and when this practice stopped can be considered a sign of business maturity. In contrast to *"fiado"* practice, only 3 SBOs exposed they had saving accounts. For example, P18 learned financial literacy with her grandmother: *"[...] I believe that it was my grandmother that told me that. She always said that we need to have money reserve. And I always like to save a little. Whenever I can, I save [...]"* indicating perceived values such as *Accountability, Identity, and Legitimization*. Those can influence microcredit access based on how they report their financial practices and business management. *"Fiado"* can also be defined as a community practice related to Identity from these SBOs, and this practice was described by SBOs as a harmful practice for their businesses, demonstrating *Accountability and Legitimization* values to improve their *Creditworthiness* towards credit agents and themselves.

*5.0.2 Accountability.* Specifically in the SBOs' social context, we identified a scenario illustrating how vulnerable people are frustrated to access a formal job. P13 explained: *"I spent 30 years of my life trying to be hired. I sent resumes, I studied and I am not getting it"* (P13). The difficulties, and frustration, to access a formal job led to entrepreneurship, not as choice, but as the only alternative to have a revenue. These coercive scenarios could be noted in other 7 SBOs'interviews. Nonetheless, we noted business *Accountability* was connected to *Human Welfare and Identity Legitimization*. For example, P16 exposed she liked to be a entrepreneur, although she was thinking to stop her business because difficulties and frustration of entrepreneurship process. The complexity of requiring documentation, submitting digital form, and planning ahead. Low-literacy to use technologies to apply for loans can make these women invisible to access microcredit. The same occurs in managing their business. They usually don't structure financial data, using informal paper annotations, giving informal credit to clients.

*5.0.3 Vulnerability. Vulnerability* in this work includes *"people who may have a low-socioeconomic status, identity as a racial and ethnic minority, those who have experienced domestic abuse, those who live in poverty or in foster care, immigrants, refugees, economically disabled, people with chronic illnesses, people who Identity as LGBTQ+, youth, children, and older adults"* [3]. Additionally, vulnerable situations are not self-excluded and can be overlaid, can generate other difficulties like low education/literacy, and can gets worse social condition. With this in mind. *Vulnerability* is related to other values such as *Universal Usability, Prejudice, Social Coercion, Identity, Frustration, and Accountability*. We identified several instances reported by five participants. The vulnerability on technology access could be perceived as part of Universal Usability value. Participants had WhatsApp access, and still had challenges to use devices and complete financial courses online (P2, P6, P15, P17, P19). Specifically, economic vulnerability like a low-traffic internet data package (P14) or low internet signal (P16) makes it difficult to access online courses. Moreover, Language skills were identified as another obstacle to immigrant people (P2). The NGO financial courses consider the *Social Accountability* to promote women as protagonists of their lives by financial empowerment and improvement of their self-esteem via *Identity Legitimization* (P2, P3, P13, P14). As an illustration, P14 highlighted *"[…] I needed to occupy my mind for not starting a depressive state, and worrying about financial, I need money […]"*. Likewise, *Frustration*, also emerged as a value when they could not access jobs in the labor market, these values were widely related with *Accountability*. More precisely seven women started on entrepreneurship due to the lack of alternatives for accessing revenue to support their families. *Vulnerabilities* can be increased in the solo mothers' context as exposed by four women. These women were submitted to a Social Coercion situation not having a legitimate choice to decide to manage a business or have a formal job in the labor market. Furthermore, when these women started their businesses there was a need to fight against prejudices of gender, age, and race (P3, P5, P7, P13, P14). For those reasons, an increase in women empowerment supported by the NGOs was crucial to foster small businesses despite the social vulnerabilities their context bring to the table.

*5.0.4 Fairness and Justice.* When debts were not paid generated name's inclusion in a in public debt list. It can be an obstacle to microcredit access. When affirmative actions are generated to promote inclusion, credit access can be considered a live change for participants. Four SBOs said that they had access microcredit when were in debt, like P4 *"I was a little bit afraid to not get a credit, because I had [credit] restriction in my name, but I did the course, and they considered it, I got the microcredit money and It's a turning point in my life"*. The sensation of financial improvement after completing the NGO financial education course was mentioned by several SBOs. Women financial inclusion perceptions were reported in nine women. For instance, P13 said *"Few people support entrepreneurship, and very few people support women entrepreneurship"*. Women financial inclusion as way to solidify *Fairness and Justice* socially. In addition, 2 women from Venezuela (P1, P18) demonstrated difficulties of communication in Portuguese. These contexts revealed the need for more development of affirmative actions that help foreign women in financial improvement.

*5.0.5 Ownership and Property. Ownership and property*, was identified as having a strong tie with *Identity. Ownership and Business Identity* were described with pride of the *Legitimization* of being part of a peripheral community, P13 exposed that *"Our suburb, for me, I think that are some of major historical places from all the cities. I came to here 9 years ago, and I am living here for 5 years. And during those 5 years I had the opportunity to know bakers, hairdressers, dread makers, singers, poets, dancers, plastic artists, graffiti artists. We lack support, we lack people opening doors for us.*

*[we] have no budget to invest in [our] dreams."* Ownership and property were also connected to *Legitimization, Coercion, Vulnerability, and Accountability*. In a vulnerable context women entrepreneur need to balance family and labor. Women submitted to social coercion to leave formal jobs to take care of their children, while at same time they need revenue (P3, P7, P20). *Collaboration* can be highlighted in a context that other people from SBOs' families collaborate working in their business (P3, P7) or a context that women had support and collaboration of neighbors, like exposed by P13 *"I started by cake business because my neighbor saw that I needed, that I knew how to do, but has not money to invest and buy supplies, so she helped me with the supplies, and I want to do the same for other women too"*.

*5.0.6 Universal Usability.* One aspect that emerged was the user-friendly technology interaction perceived by participants. The chatbot prototype was considered easy to use. When asked by the moderator about former experiences using chatbots, 6 SBOs considered that interactions were bad, like P11 *"Every day, when we call for credit card service, cell phone company, internet.I hate it!"*, or P10 *"It is very annoying, I have no patience [...]"*. Only 3 SBOs experienced good interactions with chatbots, like P5 that had problems with her card machine and solve it using the chatbot from card machine company. Additionally, P7 customized a chatbot at her WhatsApp Business app, despite not being clear what is a chatbot*"my [WhatsApp Business profile] have those automatic messages, that is the WhatsApp Business, I use it. Only cell phone things are automatic, right? When we call to cell phone company that this thing is automatic, right?"*. Moreover, *Mistrust* as a value could be perceived in relation to *Privacy*, and both in tension with *Trust*. Fear to have their data stolen by criminals as mentioned by P5 *"some text messages, sometimes you press, without wanting to and it can still your data."*, the same perspective was given by P4. Nonetheless, we think CUI mistrust could be reduced in situations likewise P15 reported:*"it has people behind developing the system, then, in this way we know that it is an automatic system, where people had to work on it."*.

*5.0.7 Identity. Identity* was associated mainly with Legitimization (P2, P3, P14, P15) in the analysis. These values are increasing women empowerment, and we noticed a tension with *Vulnerability.Identity* was tied to *Autonomy* and development of *Ownership and Property*. Women empowered her identities with the outcomes of their businesses, mainly to supply basic needs, as exposed by P3 *"it's actually helping a lot, it is giving me an income that is enough to meet my basic needs, which at the moment is what I need"*. Additionally, *Identity and Legitimization* were clustered with *Accountability, Autonomy, Human Welfare, Ownership and Property, and Universal Usability* in a context that the business is empowering women self-esteem (P14), *"Is not easy work all night, right? So then, I believe that I am a warrior. I really struggle a lot. And [I am] conqueror too, right?, because I have conquered my space, my place. It is worth it"*. *Identity* was observed in relation to *Autonomy, Ownership and Property, and Universal Usability*. Empowering women by a learning process that provide tech access, like P2 exposed *"I work with creative sewing, and tech world support us so much, right? In terms of to follow knowledge, and to share my work for other people too. The informatics, the media support. To know more people, to know other places too, this helps a lot. And in my field, for me, it is a big tool"*. Through *Identity* we see the magnitude of financial education courses to increase SBOs' self-esteem and acquire financial and tech skills to grow their businesses and empower their identities.

## 6 FINAL REMARKS

Values are not static, universal and ahistorical. They are dynamic, situated, highly dimensional, and subjective. We understand values here in relationships with people, their everyday practices, and contexts. The values we discussed in the previous sessions might have a positive, negative, or double impact, and they are tied to the power dynamics in defining, prioritizing, and evaluating different kinds of value enactments. Our analysis also identified some value

Designing multi-model conversational AI financial systems: understanding sensitive values of women entrepreneurs in Brazil   HTTF '24, June 11-12, 2024, Stockholm, Swedentensions in several of our participants' reports for instance, in *Accountability* and *Ownership and Property*. In the face of a jobless market, entrepreneurship can be a way for women to earn income. Having their own business is often tricky and may not have a constant cash flow. According to our interpretation, there is tension here, showing a problematic scenario in that women must find a way to obtain revenue for their families. Still, for them, both formal and informal job markets are complex to access. In this work, we brought to discussion essential values and value tensions to consider when designing AI systems to evaluate the business health, and amplify credit access using non-traditional data collected by conversational systems. We also provide seven recommendations based on values to be aware when designing AI conversational systems for financial services.

## REFERENCES

[1] Veronica Abebe, Gagik Amaryan, Marina Beshai, Ilene, Ali Ekin Gurgen, Wendy Ho, Naaji R. Hylton, Daniel Kim, Christy Lee, Carina Lewandowski, Katherine T. Miller, Lindsey A. Moore, Rachel Sylwester, Ethan Thai, Frelicia N. Tucker, Toussaint Webb, Dorothy Zhao, Haicheng Charles Zhao, and Janet Vertesi. 2022. Anti-Racist HCI: notes on an emerging critical technical practice. In *Extended Abstracts of the 2022 CHI Conference on Human Factors in Computing Systems (CHI EA '22)*. Association for Computing Machinery, New York, NY, USA, 1–12. https://doi.org/10.1145/3491101.3516382

[2] Ariful Islam Anik and Andrea Bunt. 2021. Data-Centric Explanations: Explaining Training Data of Machine Learning Systems to Promote Transparency. In *Proceedings of the 2021 CHI Conference on Human Factors in Computing Systems (CHI '21)*. Association for Computing Machinery, New York, NY, USA, 1–13. https://doi.org/10.1145/3411764.3445736

[3] Oghenemaro Anuyah, Karla Badillo-Urquiola, and Ronald Metoyer. 2023. Characterizing the Technology Needs of Vulnerable Populations for Participation in Research and Design by Adopting Maslow's Hierarchy of Needs. In *Proceedings of the 2023 CHI Conference on Human Factors in Computing Systems*. 1–20.

[4] A Stevie Bergman, Gavin Abercrombie, Shannon Spruit, Dirk Hovy, Emily Dinan, Y-Lan Boureau, Verena Rieser, et al. 2022. Guiding the release of safer E2E conversational AI through value sensitive design. In *Proceedings of the 23rd Annual Meeting of the Special Interest Group on Discourse and Dialogue*. Association for Computational Linguistics.

[5] Emmanuel F. Boamah, Nadine S. Murshid, and Mohammad G. N. Mozumder. 2021. A network understanding of FinTech (in)capabilities in the global South. *Applied Geography* 135 (Oct. 2021), 102538. https://doi.org/10.1016/j.apgeog.2021.102538 Publisher: Pergamon.

[6] Karen Boyd. 2022. Designing up with value-sensitive design: Building a field guide for ethical ML development. In *2022 ACM Conference on Fairness, Accountability, and Transparency*. 2069–2082.

[7] Virginia Braun and Victoria Clarke. 2006. Using thematic analysis in psychology. *Qualitative research in psychology* 3, 2 (2006), 77–101.

[8] Virginia Braun and Victoria Clarke. 2021. One size fits all? What counts as quality practice in (reflexive) thematic analysis? *Qualitative research in psychology* 18, 3 (2021), 328–352.

[9] Vitalie Bumacov, Arvind Ashta, and Pritam Singh. 2017. Credit scoring: A historic recurrence in microfinance. *Strategic Change* 26, 6 (Nov. 2017), 543–554. https://doi.org/10.1002/jsc.2165 Publisher: John Wiley & Sons, Ltd.

[10] Jo Burton. 2020. "Doing no harm" in the digital age: What the digitalization of cash means for humanitarian action. *International Review of the Red Cross* 102, 913 (April 2020), 43–73. https://doi.org/10.1017/S1816383120000491 Publisher: Cambridge University Press.

[11] Wanling Cai, Yucheng Jin, and Li Chen. 2022. Impacts of Personal Characteristics on User Trust in Conversational Recommender Systems. In *Proceedings of the 2022 CHI Conference on Human Factors in Computing Systems (CHI '22)*. Association for Computing Machinery, New York, NY, USA, 1–14. https://doi.org/10.1145/3491102.3517471

[12] Heloisa Candello, David Millen, Silvia Bianchi, Rogério de Paula, and Cláudio Pinhanez. 2015. Understanding Fiado: Informal Credit in Brazil. In *Proceedings of the 2015 Annual Symposium on Computing for Development*. 147–147.

[13] Heloisa Candello, David Millen, Claudio Pinhanez, and Silvia Bianchi. 2018. Design Insights and Opportunities from a Field Study to Digitally Enhance Microcredit Practices in Brazil. (2018).

[14] Manu Chopra, Indrani Medhi Thies, Joyojeet Pal, Colin Scott, William Thies, and Vivek Seshadri. 2019. Exploring Crowdsourced Work in Low-Resource Settings. In *Proceedings of the 2019 CHI Conference on Human Factors in Computing Systems (CHI '19)*. Association for Computing Machinery, New York, NY, USA, 1–13. https://doi.org/10.1145/3290605.3300611

[15] Victoria Clarke and Virginia Braun. 2021. Thematic analysis: a practical guide. *Thematic Analysis* (2021), 1–100.

[16] Eric Corbett and Christopher Le Dantec. 2021. Designing Civic Technology with Trust. In *Proceedings of the 2021 CHI Conference on Human Factors in Computing Systems (CHI '21)*. Association for Computing Machinery, New York, NY, USA, 1–17. https://doi.org/10.1145/3411764.3445341

[17] John W Creswell and J Creswell. 2003. *Research design*. Sage publications Thousand Oaks, CA.

[18] Upol Ehsan, Philipp Wintersberger, Q. Vera Liao, Martina Mara, Marc Streit, Sandra Wachter, Andreas Riener, and Mark O. Riedl. 2021. Operationalizing Human-Centered Perspectives in Explainable AI. In *Extended Abstracts of the 2021 CHI Conference on Human Factors in Computing Systems (CHI EA '21)*. Association for Computing Machinery, New York, NY, USA, 1–6. https://doi.org/10.1145/3411763.34413427

[19] Upol Ehsan, Philipp Wintersberger, Q. Vera Liao, Elizabeth Anne Watkins, Carina Manger, Hal Daumé III, Andreas Riener, and Mark O Riedl. 2022. Human-Centered Explainable AI (HCXAI): Beyond Opening the Black-Box of AI. In *Extended Abstracts of the 2022 CHI Conference on Human Factors in Computing Systems (CHI EA '22)*. Association for Computing Machinery, New York, NY, USA, 1–7. https://doi.org/10.1145/3491101.3503727

[20] Michael Evans, Lianne Kerlin, Joanne Parkes, and Todd Burlington. 2022. "I want to be independent. I want to make informed choices.": An Exploratory Interview Study of the Effects of Personalisation of Digital Media Services on the Fulfilment of Human Values. In *Proceedings of the 2022 ACM International Conference on Interactive Media Experiences* (Aveiro, JB, Portugal) *(IMX '22)*. Association for Computing Machinery, New York, NY, USA, 325–330. https://doi.org/10.1145/3505284.3532977

[21] Sina Fazelpour and Maria De-Arteaga. 2022. Diversity in sociotechnical machine learning systems. *Big Data & Society* 9, 1 (2022), 20539517221082027.

[22] Batya Friedman and David G Hendry. 2019. *Value sensitive design: Shaping technology with moral imagination*. Mit Press.

[23] Batya Friedman, Peter H Kahn, Alan Borning, and Alina Huldtgren. 2013. Value sensitive design and information systems. *Early engagement and new technologies: Opening up the laboratory* (2013), 55–95.

[24] Daniel Gregory and Diego Monteiro. 2023. Is this the real life? Investigating the credibility of synthesized faces and voices created by amateurs using artificial intelligence tools.. In *Proceedings of the 2023 ACM International Conference on Interactive Media Experiences Workshops* (, Nantes, France,) *(IMXw '23)*. Association for Computing Machinery, New York, NY, USA, 118–122. https://doi.org/10.1145/3604321.3604329

[25] Samia Ibtasam, Lubna Razaq, Haider W. Anwar, Hamid Mehmood, Kushal Shah, Jennifer Webster, Neha Kumar, and Richard Anderson. 2018. Knowledge, Access, and Decision-Making: Women's Financial Inclusion In Pakistan. In *Proceedings of the 1st ACM SIGCAS Conference on Computing and Sustainable Societies (COMPASS '18)*. Association for Computing Machinery, New York, NY, USA, 1–12. https://doi.org/10.1145/3209811.3209819

[26] Pranjal Jain, Alex Jordan Blandin, Jacki O'Neill, Mark Perry, Samia Ibtasam, Suleman Shahid, Beni Chugh, David Sullivan, Heloisa Candello, James Pomeroy, Rajat Jain, Robert Dowd, Matt Roach, and Matt Jones. 2022. Platformisation of Digital Financial Services (DFS): The Journey of DFS in the Global North and Global South. In *Extended Abstracts of the 2022 CHI Conference on Human Factors in Computing Systems (CHI EA '22)*. Association for Computing Machinery, New York, NY, USA, 1–5. https://doi.org/10.1145/3491101.3516507

[27] Pranjal Jain, Rama Adithya Varanasi, and Nicola Dell. 2021. "Who is protecting us? No one!" Vulnerabilities Experienced by Low-Income Indian Merchants Using Digital Payments. In *ACM SIGCAS Conference on Computing and Sustainable Societies (COMPASS '21)*. Association for Computing Machinery, New York, NY, USA, 261–274. https://doi.org/10.1145/3460112.3471961

[28] Joseph Jofish Kaye, Mary McCuistion, Rebecca Gulotta, and David A Shamma. 2014. Money talks: tracking personal finances. In *Proceedings of the SIGCHI Conference on Human Factors in Computing Systems*. 521–530.

[29] Raina Langevin, Ross J Lordon, Thi Avrahami, Benjamin R. Cowan, Tad Hirsch, and Gary Hsieh. 2021. Heuristic Evaluation of Conversational Agents. In *Proceedings of the 2021 CHI Conference on Human Factors in Computing Systems (CHI '21)*. Association for Computing Machinery, New York, NY, USA, 1–15. https://doi.org/10.1145/3411764.3445312

[30] Minha Lee, Jaisie Sin, Guy Laban, Matthias Kraus, Leigh Clark, Martin Porcheron, Benjamin R. Cowan, Asbjørn Følstad, Cosmin Munteanu, and Heloisa Candello. 2022. Ethics of Conversational User Interfaces. In *Extended Abstracts of the 2022 CHI Conference on Human Factors in Computing Systems (CHI EA '22)*. Association for Computing Machinery, New York, NY, USA, 1–7. https://doi.org/10.1145/3491101.3503699

[31] Nora McDonald, Sarita Schoenebeck, and Andrea Forte. 2019. Reliability and inter-rater reliability in qualitative research: Norms and guidelines for CSCW and HCI practice. *Proceedings of the ACM on human-computer interaction* 3, CSCW (2019), 1–23.

[32] David Mhlanga. 2020. Industry 4.0 in Finance: The Impact of Artificial Intelligence (AI) on Digital Financial Inclusion. *International Journal of Financial Studies* 8, 3 (Sept. 2020), 45. https://doi.org/10.3390/ijfs8030045 Number: 3 Publisher: Multidisciplinary Digital Publishing Institute.

[33] Aki Nagano. 2018. Economic Growth and Automation Risks in Developing Countries Due to the Transition Toward Digital Modernity. In *Proceedings of the 11th International Conference on Theory and Practice of Electronic Governance (ICEGOV '18)*. Association for Computing Machinery, New York, NY, USA, 42–50. https://doi.org/10.1145/3209415.3209442

[34] Adasa Nkrumah, Li Ping, Anjum Safia., and Md Altab Hossin. 2018. Mobile Money Transfer: The Process Model Perspective. In *Proceedings of the 2018 9th International Conference on E-business, Management and Economics (ICEME 2018)*. Association for Computing Machinery, New York, NY, USA, 28–35. https://doi.org/10.1145/3271972.3271986

[35] Craig Plain. 2007. Build an affinity for KJ method. *Quality Progress* 40, 3 (2007), 88.

[36] Upul Anuradha Rathnayake, Thilina Halloluwa, Pradeepa Bandara, Medhani Narasinghe, and Dhaval Vyas. 2021. Exploring Entrepreneurial Activities in Marginalized Widows: A Case from Rural Sri Lanka. *Proc. ACM Hum.-Comput. Interact.* 5, CSCW1 (April 2021), 142:1–142:24. https://doi.org/10.1145/3449216

[37] Brianna Richardson, Jean Garcia-Gathright, Samuel F. Way, Jennifer Thom, and Henriette Cramer. 2021. Towards Fairness in Practice: A Practitioner-Oriented Rubric for Evaluating Fair ML Toolkits. In *Proceedings of the 2021 CHI Conference on Human Factors in Computing Systems (CHI '21)*. Association for Computing Machinery, New York, NY, USA, 1–13. https://doi.org/10.1145/3411764.3445604

[38] Arne Rieber, Frank Bliss, and Karin Gaesing. 2022. Sustainable Financial Inclusion in the Rural Area. *Institute for Development and Peace (INEF), University of Duisburg-Essen* 26 (2022). https://www.researchgate.net/publication/361668354_Sustainable_Financial_Inclusion_in_the_Rural_Area

[39] Alexandra Rizzi, Alexandra Kessler, and Jacobo Menajovsky. 2021. The Stories Algorithms Tell: Bias and Financial Inclusion at the Data Margins. *Center for Financial Inclusion, Accion* (March 2021), 35. https://content.centerforfinancialinclusion.org/wp-content/uploads/sites/2/2021/03/The-Stories-Algorithms-Tell-CFI-publication-MAR21.pdf

[40] Yasaman Rohanifar, Sharifa Sultana, Swapnil Nandy, Pratyasha Saha, Md. Jonayed Hossain Chowdhury, Mahdi Nasrullah Al-Ameen, and Syed Ishtiaque Ahmed. 2022. The Role of Intermediaries, Terrorist Assemblage, and Re-skilling in the Adoption of Cashless Transaction Systems in Bangladesh.

Designing multi-model conversational AI financial systems: understanding sensitive values of women entrepreneurs in Brazil

In *ACM SIGCAS/SIGCHI Conference on Computing and Sustainable Societies (COMPASS) (COMPASS '22)*. Association for Computing Machinery, New York, NY, USA, 266–279. https://doi.org/10.1145/3530190.3534810

[41] Neelima Sailaja, Rhianne Jones, and Derek McAuley. 2021. Human Data Interaction in Data-Driven Media Experiences: An Exploration of Data Sensitive Responses to the Socio-Technical Challenges of Personal Data Leverage. In *Proceedings of the 2021 ACM International Conference on Interactive Media Experiences* (Virtual Event, USA) *(IMX '21)*. Association for Computing Machinery, New York, NY, USA, 108–119. https://doi.org/10.1145/3452918.3458797

[42] Rahime Belen Sağlam and Jason R. C. Nurse. 2020. Is your chatbot GDPR compliant? Open issues in agent design. In *Proceedings of the 2nd Conference on Conversational User Interfaces (CUI '20)*. Association for Computing Machinery, New York, NY, USA, 1–3. https://doi.org/10.1145/3405755.3406131

[43] Anna Sheremetieva, Ihor Romanovych, Sam Frish, Mykola Maksymenko, and Orestis Georgiou. 2023. What's my future: a Multisensory and Multimodal Digital Human Agent Interactive Experience. In *Proceedings of the 2023 ACM International Conference on Interactive Media Experiences* (Nantes, France) *(IMX '23)*. Association for Computing Machinery, New York, NY, USA, 40–46. https://doi.org/10.1145/3573381.3596161

[44] Roger Tafotie. 2020. Fostering Digital Financial Services in Africa: A Case of Embracing Innovation for Business and Inclusion. https://doi.org/10.2139/ssrn.3557808

[45] Takane Ueno, Yuto Sawa, Yeongdae Kim, Jacqueline Urakami, Hiroki Oura, and Katie Seaborn. 2022. Trust in Human-AI Interaction: Scoping Out Models, Measures, and Methods. In *Extended Abstracts of the 2022 CHI Conference on Human Factors in Computing Systems (CHI EA '22)*. Association for Computing Machinery, New York, NY, USA, 1–7. https://doi.org/10.1145/3491101.3519772

[46] Richmond Y. Wong. 2021. Using Design Fiction Memos to Analyze UX Professionals' Values Work Practices: A Case Study Bridging Ethnographic and Design Futuring Methods. In *Proceedings of the 2021 CHI Conference on Human Factors in Computing Systems (CHI '21)*. Association for Computing Machinery, New York, NY, USA, 1–18. https://doi.org/10.1145/3411764.3445709

[47] Richmond Y. Wong and Tonya Nguyen. 2021. Timelines: A World-Building Activity for Values Advocacy. In *Proceedings of the 2021 CHI Conference on Human Factors in Computing Systems (CHI '21)*. Association for Computing Machinery, New York, NY, USA, 1–15. https://doi.org/10.1145/3411764.3445447

[48] Sarah Yu and Samia Ibtasam. 2018. A Qualitative Exploration of Mobile Money in Ghana. In *Proceedings of the 1st ACM SIGCAS Conference on Computing and Sustainable Societies (COMPASS '18)*. Association for Computing Machinery, New York, NY, USA, 1–10. https://doi.org/10.1145/3209811.3209863

[49] Qingxiao Zheng, Yiliu Tang, Yiren Liu, Weizi Liu, and Yun Huang. 2022. UX Research on Conversational Human-AI Interaction: A Literature Review of the ACM Digital Library. In *Proceedings of the 2022 CHI Conference on Human Factors in Computing Systems (CHI '22)*. Association for Computing Machinery, New York, NY, USA, 1–24. https://doi.org/10.1145/3491102.3501855

[50] Xinyu Zhu, Xingguo Zhang, Zinan Chen, Zhanxun Dong, Zhenyu Gu, and Danni Chang. 2022. The Trusted Listener: The Influence of Anthropomorphic Eye Design of Social Robots on User's Perception of Trustworthiness. In *Proceedings of the 2022 CHI Conference on Human Factors in Computing Systems (CHI '22)*. Association for Computing Machinery, New York, NY, USA, 1–13. https://doi.org/10.1145/3491102.3517670

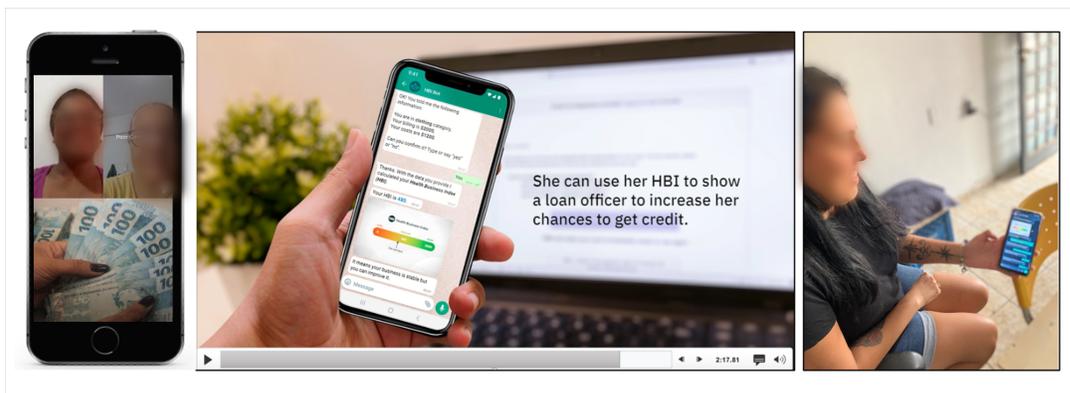

Fig. 1. Online semi-structured interviews, and Cycle 2 and 3 in-person semi-structured interviews. In both settings, small business owners gave their view of the conversational system. The video-prototype in Cycle 1, and the interactive pilot in Cycle 2 were used as a probe to discuss the main the perceived inherent values, and value tensions built in their everyday practices

Table 1. Participants and Business activities

| Part. | Interview cycle | Business activity |
|---|---|---|
| P1 | 1 | candy/cake |
| P2 | 1 | handbag/wallet/sewn accessories |
| P3 | 1 | candy/cake |
| P4 | 1 | candy/cake/clothes |
| P5 | 1 | craftwork |
| P6 | 1 | perfumery/cosmetics |
| P7 | 1 | candy/cake |
| P8 | 1 | hair services |
| P9 | 1 | personalized gifts |
| P10 | 1 | cake/nail service |
| P11 | 1 | candy/cake |
| P12 | 1 | hair services |
| P13 | 2 | candy/cake |
| P14 | 2 | scarf craftwork |
| P15 | 2 | cake decoration |
| P16 | 2 | clothes/accessories |
| P17 | 3 | crochet/ cleaning service |
| P18 | 3 | macrame/ craftwork |
| P19 | 3 | crochet/ craftwork |
| P20 | 3 | candy/cake/craftwork |

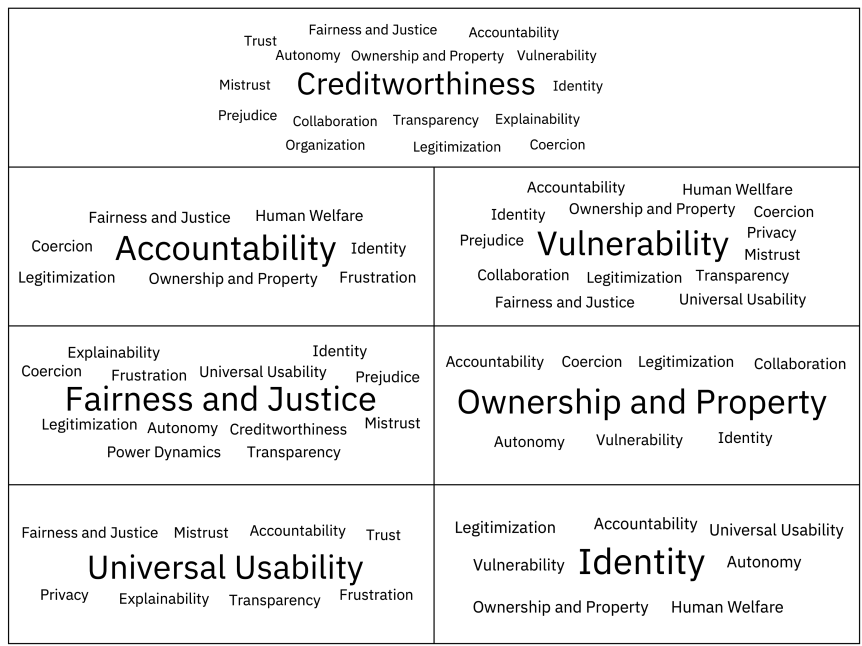

Fig. 2. Affinity Diagram for values, including Global Themes and basic themes

| Theme/Value | Description | Design Recommendation |
| --- | --- | --- |
| Creditworthiness | This expresses the limit in that a particular individual is considered able to receive financial credit. | [R01]: *Creditworthiness* should be considered a complex metric comprising formal data, social capital practices, business structure/project, level of Accountability, vulnerability status, tech literacy, and other social-contextual elements. A conversational AI system should cover those topics to have a comprehensive picture of SBOs and businesses context. |
| Accountability | This involves the responsibility- owner and the level that a person or group can to answer for an action. | [R02]: *Accountability* as an evaluation criteria, should be considered including other factors as: SBOs low-education, low tech-literacy, financial reality and context. CUIs should evaluate their context and guide them throughout the process and actions required for loan acquisitions. |
| Vulnerability | This designates situations to which a person or group is subject socially and that can harm you, whether by prejudice or by particularities that can block/complicate access to conditions of subsistence/quality of life. | [R03]: *Vulnerability* should be considered in terms of SBOs social profile, understanding different vulnerabilities in their lives, that can create specific obstacles to empower their business and lives and increase ways to microcredit access. |
| Fairness and Justice | This is related to perception and actions that can improve SBOs lives according their particularities. | [R04]: *Fairness and Justice* should be considered when designing conversational systems. Users should be able to communicate in their language, considering the diversity of people in low-income situations that apply for loans. |
| Ownership and Property | This aggregates SBOs' feelings and their perceptions about their business, ways to entrepreneur, and possible improvement in their businesses. | [R05]: *Ownership and Property* can be covered in conversational systems asking not only about formal legitimization of their business, but also women support network should be considered. |
| Universal Usability | This relates to literacy for interaction with tech and, specifically in this context, with CUI and AI. | [R06]: *Universal Usability* should be considered when there are different levels of tech literacy, avoid confusing answers and provide clear tech explanations to increase trust. |
| Identity | This assemble characteristics from person and their business that are intrinsically related to their values, ideas, and self construction. | [R07]: *Identity* should be considered as the most sensitive value, and the system needs to consider SBO personal context to ask questions and suggestions using a vocabulary that can boost SBO self-confidence and development. |

Table 2. Description and Design Recommendations according Global Themes.